\documentclass{iau} 
\usepackage{graphicx}
\usepackage{natbib}

\title[Galactic magnetic fields of a stellar origin]{A supernova scenario for magnetic fields and rotation measures in galaxies}

\author[A.M. Beck \& K. Dolag]{Alexander M. Beck \& Klaus Dolag}

\affiliation{University Observatory Munich \\ Scheinerstr. 1, D-81679 Munich, Germany \\email: {\tt abeck@usm.lmu.de}}

\pubyear{2015}
\volume{315}  
\jname{From interstellar clouds to star-forming galaxies: universal processes?}
\editors{P. Jablonka, F. van der Tak \& P. Andr\`{e}, eds.}
\begin{document}

\maketitle

\begin{abstract}
We present a model for the seeding and evolution of magnetic fields in galaxies by supernovae (SN).
SN explosions during galaxy assembly provide seed fields, which are subsequently amplified by compression, shear flows and random motions.
Our model explains the origin of $\mu$G magnetic fields within galactic structures.
We implement our model in the MHD version of the cosmological simulation code Gadget-3 and couple it with a multi-phase description of the interstellar medium.
We perform simulations of Milky Way-like galactic halo formation and analyze the distribution and strength of the magnetic field.
We investigate the intrinsic rotation measure (RM) evolution and find RM values exceeding 1000 rad/m$^2$ at high redshifts and RM values around 10 rad/m$^2$ at present-day.
We compare our simulations to a limited set of observational data points and find encouraging similarities.
In our model, galactic magnetic fields are a natural consequence of the very basic processes of star formation and galaxy assembly.
\end{abstract}

\section*{The simulations}

\begin{figure}[b]
\begin{center}
\includegraphics[angle=90, width=0.9\textwidth]{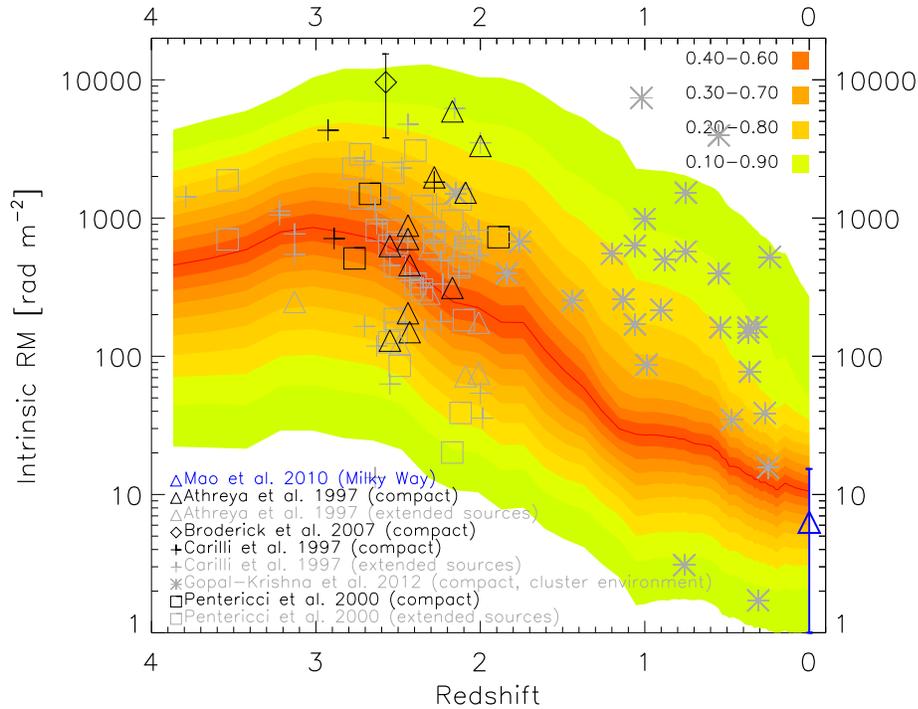} 
\caption{Mean intrinsic RM of the simulated galactic halo compared to data points taken from the literature.
The black data points are suited for comparison to our simulation (colors), while the grey data points represent extended source sizes or denser environments.}
\label{sim_RM}
\end{center}
\end{figure}

\noindent{}We extend the studies of \cite{beck12} and \cite{beck13} and follow the cosmological evolution of a Milky Way-like halo from high redshifts until z=0.
Coupled to the star formation process, we inject small-scale dipole-shaped magnetic seed fields \citep[see also][]{rees87} given typical dimensions and magnetic field amplitudes found in canonical SN remnants.
The seeding rate roughly corresponds to 10$^{-9}$ G/Gyr.
This is the only process of magnetic field seeding and we do not consider any other seeds as in many previous simulations \citep[see also][]{beck12}.
Subsequently, the magnetic field strength increases exponentially on timescales of a few ten million years within the innermost regions of the halo.
The amplification is due to dynamo action occurring during galaxy formation and at last, motions and turbulent diffusion carry the field towards the halo outskirts and magnetize the entire halo at z=0.
The simulations were run with the Gadget \citep{springel05} code and the \cite{springel03} star formation prescription.

\section*{Rotation measure and magnetic field evolution}

\noindent{}Figure \ref{sim_RM} shows the intrinsic RM evolution of the innermost region of the halo and observational data taken from the literature.
The $\mathrm{RM}\sim{}\int_\ell{}{B_{||}n_e}d\ell{}$ is calculated as the line-of-sight integral of magnetic field and thermal electron density.
In our simulations, at redshifts between z=4 and z=2, high intrinsic RM values exceeding 1000 rad/m$^2$ are due to high star formation rates and very efficient field amplification.
As the halo virializes towards z=0, the magnetic field diffuses and therefore the intrinsic RM values decline to 10 rad/m$^2$.
At the moment, observational RM data at high redshifts is difficult to obtain and interpret.
However, we think a careful comparison with existing data is already possible.
These data shows both trends reproduced by our simulations for objects residing in the same environments and similar source sizes.
Galaxy formation at high redshifts undergoes extremely violent phases with excessive amounts of feedback.
This could be the origin of the large intrinsic RM values see at that cosmic epoch.

\section*{Acknowledgments}
\noindent{}This work is supported by the DFG Cluster of Excellence 'Universe', DFG Research Unit '1254' and by the 'Magneticum' project (http://www.magneticum.org).


\begin{thebibliography}{}
\bibitem[\protect\citeauthoryear{Athreya et al.}{1998}]%
  {athreya98} Athreya~R.M., et al., 1998, \textit{A\&A}, 329, 809 

\bibitem[\protect\citeauthoryear{Beck et al.}{2012}]%
  {beck12} Beck~A.M., et al., 2012, \textit{MNRAS}, 422, 2152

\bibitem[\protect\citeauthoryear{Beck et al.}{2013}]%
  {beck13} Beck~A.M., et al., 2013, \textit{MNRAS}, 435, 3575

\bibitem[\protect\citeauthoryear{Broderick et al.}{2007}]%
  {broderick07} Broderick~J.W., et al., 2007, \textit{MNRAS}, 375, 1059 

\bibitem[\protect\citeauthoryear{Carilli et al.}{1997}]%
  {carilli97} Carilli~C.L., et al., 1997, \textit{ApJS}, 109, 1 

\bibitem[\protect\citeauthoryear{Gopal-Krishna et al.}{2012}]%
  {gopal12} Gopal-Krishna, et al., 2012, \textit{ApJ}, 744, 31 

\bibitem[\protect\citeauthoryear{Mao et al.}{2010}]%
  {mao10} Mao~S.A., et al., 2010, \textit{ApJ}, 714, 1170 

\bibitem[\protect\citeauthoryear{Pentericci et al.}{2000}]%
  {pentericci00} Pentericci~L., et al., 2000, \textit{A\&AS}, 145, 121 

\bibitem[\protect\citeauthoryear{Rees}{1987}]%
  {rees87} Rees~M.J., 1987, \textit{QJRAS}, 28, 197 
  
\bibitem[\protect\citeauthoryear{Springel \& Hernquist}{2003}]%
  {springel03} Springel~V. \& Hernquist~L., 2003, \textit{MNRAS}, 339, 289

\bibitem[\protect\citeauthoryear{Springel}{2005}]%
  {springel05} Springel~V., 2005, \textit{MNRAS}, 364, 1105
\end{thebibliography}
\end{document}